\documentclass[aps,pra,10pt,twocolumn,showpacs,superscriptaddress,floatfix,footinbib,subfigure]{revtex4}
\usepackage{amsmath}
\usepackage{latexsym}
\usepackage{amssymb}
\usepackage{graphics,epstopdf,subfigure,color,hyperref}
\definecolor{Red}{rgb}{1,0,0}

{\catcode`\|=\active
  \gdef\Braket#1{\begingroup
\mathcode`\|32768\let|\BraVert\left<{#1}\right>\endgroup}}
\def\BraVert{\egroup\,\mid\,\bgroup}

\newtheorem{property}{Property}

\begin{document}

{\bf Comment on `How the result of a single coin toss can turn out to be 100 heads'} \hfill\break

In \cite{FC} the authors claim that `weak values are not inherently quantum but rather a purely statistical feature'. I argue that their model reproduces only few elements of weak measurements but fails to reproduce all the other intrinsically quantum features.  

The quantum nature of post-selection paradoxes and weak values has been discussed previously in the literature \cite{Kirkpatrick2003,RV,Leifer2005,Danan2013, Vaidman2013} with  examples of classical systems that supposedly exhibit many of the strange properties related to post-selection\cite{Tomita,Dressel}. The fact that many weak measurement experiments are often  performed with classical light is well known and was also addressed in the past \cite{Danan2013,Bliokh,Dennis,Vaidman2013}. To that matter, the paper is neither novel nor comprehensive enough. However, I will adopt the constructive approach of the recent \cite{Pusey,Ipsen} focusing now only on some missing elements in the analysis of FC.

The following arguments are general, but some will be best illustrated using the `least invasive' case where the so called {\it disturbance parameter} $\delta$ achieves its maximal value $1-\delta=\lambda$, where $\lambda$ denotes the measurement strength. This bound gives the lowest probability for an orthogonal post-selection without inducing  negative probabilities in Eqs. 25 and 27 in \cite{FC} thus making the suggested scheme as close as possible to its quantum counterpart. In this case, Eq. 27 reads $Pr(\phi=-1|s=-1,\psi=+1)=0$. Hence, post-selecting $-1$ is only consistent with an outcome of $s=+1$. This observation explains  the `anomalous' weak value obtained in Eq. 33.

The first inconsistency with quantum weak measurements \cite{AAV} appears when the pre- and post-selection coincide, that is, when the expectation value is calculated. By Eq. 26 the outcome will be $s=\pm1$ with equal probability (the probability is $s$-independent) yielding the erroneous weak value $a_w=0$. This is in contrast to the quantum scheme of weak measurement which has the following trivial property:

\begin{property} In the case where the pre- and post-selection coincide, the weak value equals the expectation value \cite{AAV}.
\end{property}

Thus yielding the weak value $a_w=+1$, when $|\psi\rangle=|\phi\rangle=|+1\rangle$.

Property 1 can be viewed as a direct consequence of the non-invasive nature of weak measurements, another property that is missing in the FC example. At the `least invasive' limit it is evident that  $a_w=\infty$. Presumably this is in line with the formula for calculating quantum weak values for orthogonal pre- and post-selection. However, this scenario is forbidden by quantum weak measurements due to the following property:

\begin{property} Weak measurements are non-invasive: Following a weak measurement the probability of post-selecting an orthogonal state is vanishingly small \cite{AAV}.\end{property}

If this were true in the FC example, the main result (Eq. 33) would have never been observed. In more precise terms, the probability of post-selecting an orthogonal state is of second order in the weakness parameter, while in the FC setup it is of first order.  Unlike the simplistic FC setup, post-selection for a weak measurement does not require  invasive measurements. It is possible to gain information about a pre- and post-selected ensemble (and in particular, to measure an `anomalous' weak value) via a quantum weak measurements without causing any particle to change its state to an orthogonal state. Non-invasiveness has a further consequence:

\begin{property} Given enough pre- and post-selected systems that have the same weak value for a given observable, it is possible to measure the weak value (in a non-invasive way) to an arbitrary precision \cite{Tollaksen2007}.\end{property}

This property is not only a consequence of non-invasiveness, but also a consequence of coherence. While lack of coherence is an obvious problem with any classical model, the FC example completely avoids the effect of the weak interaction on the measurement device. However this effect is probably the most important property of quantum weak measurements:

\begin{property} In the weak measurement formalism there is a coherent interaction. The measurement pointer is affected by an effective interaction Hamiltonian proportional to the weak value \cite{potential}.
\end{property} 
It is this property that gives weak values their physical significance.



In light of the above, we see that some important properties of quantum weak measurements are not captured by the FC example.  Therefore, while the ongoing debate on the conceptual and experimental virtues of weak measurements is not yet settled, the FC setup adds nothing above the arguments already known in literature. Moreover, it can be misleading with regard to the quantum nature of weak values in the general case. \hfill\break




Eliahu Cohen \hfill\break
School of Physics and Astronomy, Tel Aviv University, Tel Aviv 6997801, Israel \hfill\break

\end{document}